\documentclass[10pt,conference]{IEEEtran}
\usepackage{graphicx}
\usepackage{flushend}
\usepackage{todonotes}

\begin{document}
\title{Learning Software Quality Assurance with Bricks}

\author{\IEEEauthorblockN{Miguel Eh\'ecatl Morales-Trujillo}
\IEEEauthorblockA{University of Canterbury\\
Christchurch, New Zealand\\
miguel.morales@canterbury.ac.nz}}

\maketitle

\begin{abstract}
Software Quality Assurance (SQA) and Software Process Improvement (SPI) are topics of crucial importance for software engineers; however, teaching them in a lecture room comes with several limitations due to lack of practical experience. 
With that in mind, we created KUALI-Brick, a LEGO\textsuperscript{\textregistered}-based activity that brings SQA and SPI concepts together applying them in order to successfully build a LEGO city.
This hands-on activity has been carried out in a fourth-year Software Engineering course at the University of Canterbury, with current results showing high levels of fun, increased engagement and an improved learning experience. 
We present a step-by-step guide to replicate the activity as well as lessons learned after conducting the activity for three consecutive years.
\end{abstract}

\IEEEpeerreviewmaketitle

\section{Introduction}
Over the years, the term quality has been defined from various perspectives. For example, it has been seen as the ``conformance to requirements'' \cite{crosby1979} for ``achieving excellent levels of fitness for use'' \cite{humphrey1989} where ``the customer is the final arbiter and it is market-driven'' \cite{kan2003}.

Nonetheless, in the words of Pfleeger and Kitchenham, ``quality is hard to define, impossible to measure, easy to recognise'' \cite{pfleeger1996}.
According to Gillies, ``quality is generally transparent when present, but easily recognized in its absence'' \cite{gillies2011}. 
Based on these definitions, it is evident that quality, although purely abstract, must be instantiated in some way in order to be demonstrated. 

In the context of Software Engineering, the term ``software quality'' is defined as ``the “capability of a software product to satisfy stated and implied needs under specified conditions'' \cite{isoiec25010}. 
Software quality is ``the degree to which a system, component, or process meets (i) specified requirements, and (ii) customer or user needs or expectations'' \cite{ieee730}.

The concept of software quality is crucial for software engineers, who have to ensure that software development processes they follow together with software products they create are of high quality.
In other words, software engineers have no choice but to show that their processes and products are suitable and conform to the specified requirements and expectations.
In order to prove this suitability and compliance, Software Quality Assurance (SQA) and Software Process Improvement (SPI) play a fundamental role in software quality.

While SQA is defined as ``a planned and systematic pattern of all actions necessary to provide adequate confidence that an item or product conforms to established technical requirements'' \cite{ieee610}, SPI represents ``a planned, managed and controlled effort which aims to enhance the capability of the software development processes of an organization'' \cite{baldassarre2009}.

Both, SQA and SPI, cannot be taught by applying a single standard approach or a single set of common procedures as students need hands-on experience with practical scenarios \cite{kurkovsky2019}. 
Following the motivation of providing Software Engineering students with hands-on practice in the classroom, in this experience report we present KUALI\footnote{KUALI: Nahuatl word meaning good, fine or appropriate.}-Brick: a LEGO\textsuperscript{\textregistered}-based activity that aims at understanding and applying SQA and SPI concepts and techniques while participants build objects and assess their quality based on metrics associated to specific quality attributes.

On the one hand, we have chosen LEGO as it offers the means to make quality concepts tangible. On the other hand, activities that involve playing in the classroom offer several important benefits in order
to enhance the learning process~\cite{kurkovsky2019}.

We also present some preliminary empirical results and lessons learned after having successfully performed KUALI-Brick activity for the last three years with fourth-year Software Engineering students in a SQA course at the University of Canterbury, New Zealand. 

The remainder of this paper is structured as follows: Section \ref{sec:related_work} describes current work on the use of LEGO in Software Engineering education. 
Section \ref{sec:seng403} and \ref{sec:kuali-brick} expand on the SQA course description and the KUALI-Brick activity respectively. 
Section \ref{sec:discussion} consolidates discussion of the results. Finally, conclusions and future work are presented in Section \ref{sec:conclusions}.

\section{Related Work}
\label{sec:related_work}
In this section we present results of a literature review on the topic as well as a summary of the papers found.

\subsection{Literature Review}
While LEGO popularity is indisputable in the toy industry, its usage has definitely crossed boundaries to other fields with the education sector being one of the most striking examples. 
In the context of this paper, our interest is focused  on identifying those educational activities that use LEGO in Software Engineering.
This motivated the following research question:

\begin{center} 
\textit{How is LEGO used in Software Engineering education?}
\end{center}

To find answers to the aforementioned question, we performed a literature review with a search run in four databases: Scopus, IEEE Xplore, ACM Digital Library and Wiley Online Library.
The search string was the following:

\begin{center} 
``lego AND education''
\end{center}

The search scope was focused on peer-reviewed research papers published in journals, conferences and workshops. 
The inclusion criteria applied for the selection procedure were:
\begin{itemize}
    \item Papers that satisfy the search string.
    \item Papers published in journals, conferences or workshops.
    \item Papers written in English.
\end{itemize}
The exclusion criteria considered were:
\begin{itemize}
    \item Papers with no focus on Software Engineering education.
    \item Papers dealing with programming or robotics.
\end{itemize}

The last exclusion criteria turned out to be necessary after obtaining a considerably high number of papers that address LEGO Mindstorms in the context of programming and robotics. While these topics are related to Software Engineering, they are not considered to be within the scope of this research.

The search was executed on October 1\textsuperscript{st}, 2020. The total number of returned papers was 776, with 241 from ACM Digital Library, 435 from Scopus, 92 from IEEE Xplore and 8 from Wiley Online Library.
After applying the inclusion and exclusion criteria, 16 papers were identified as relevant (see Table \ref{tab:primaries}).

For the purposes of an accurate data extraction, a template containing fields for paper ID, authors, title, year of publication, document type, abstract, topic, data collection method applied, number of participants, and results was created. 
The findings of the literature review are presented in the following subsection.

\subsection{Software Engineering Education with LEGO}
Teaching Agile frameworks, particularly Scrum and its related concepts, is the most recurrent topic in which LEGO is being used.
Lynch et al. created a boot camp for learning Agile software engineering concepts with LEGO bricks as the medium \cite{lynch2011}.
Authors reported that most of the students enjoyed the boot camp, however, their apparent retention of the information was not significantly better than any of the other course information \cite{lynch2011}.

Bica and da Silva \cite{bica2020} designed a classroom activity to teach Scrum concepts to undergraduate students.
The authors reported that the knowledge of the students in regards to Scrum increased, same as their motivation.

Stegh\"{o}fer et al. reported teaching Scrum with LEGO as a means of providing practical experience to anchor theoretical knowledge \cite{steghofer2016}.
Their results showed that most students were positive about the activity and felt pleased with the hands-on experience.

Paasivaara et al. created a LEGO-based simulation game to support the students' learning of roles, events and concepts in Scrum \cite{paasivaara2014}.
The simulation game was evaluated with a survey and student lecture diaries.
The authors concluded that the simulation game was a valuable tool for teaching Scrum to university students, the students' reaction to the game was positive, and the learning outcome improved \cite{paasivaara2014}.

LEGO bricks have also been used to introduce students and IT professionals into Agile and XP concepts \cite{lubke2005}, the main ideas and practices of TDD \cite{kurkovsky2016} and the use of Essence \cite{steghofer2019}, a standard that aims at defining and consequently improving Software Engineering practices.  

In \cite{steghofer2018}, LEGO is used to introduce the concepts of SPI and Scrum into a Bachelor level course with the aim of providing students with a direct experience of observing issues, creating an improvement plan to address them, and finally, applying and evaluating the plan. 
Anecdotally, the KUALI-Brick activity presented in this paper originated when the author of the paper was teaching Kanban and Scrum to IT professionals in 2013, having evolved over the years.

In addition, LEGO Serious Play\textsuperscript{TM} (LSP) has been used to teach several core software engineering topics through hands-on scenarios, such as requirements engineering \cite{kurkovsky2019} \cite{kurkovsky2015} \cite{mayr2018} and risk identification \cite{kurkovsky2015}. 
Initial results suggest that LSP has a positive impact on student learning, while also increasing student engagement with the course material \cite{kurkovsky2015}. 
However, no significant improvement has been identified \cite{kurkovsky2019}.

Gama \cite{gama2019} reported an active learning approach where LEGO bricks are used to exemplify mini-project scenarios with the objective to teach requirements engineering and project planning.

Teaching particular strategies or learning specific skills are other goals that have been pursued by means of LEGO.
For example, a bridge building project has been reported in \cite{hood2006}; the project teaches students how to measure progress and manage change. Other examples are: allowing students to practice collaborative design, parallel development, and component integration \cite{kurkovsky2018}.

\v{S}ablis et al. used LEGO bricks to help students understand the communication and coordination challenges, and problems that might arise in global software engineering (GSE) projects, with positive results \cite{sablis2019}.

Lastly, Laird and Yang presented a method that uses LEGO to allow students to get hands-on practice in their estimation skills and techniques as well as in testing some of the commonly held estimation heuristics \cite{laird2016}.

To sum up, the primary papers reported studies involving from 15 to 200 participants. 
While surveys are the most common means of validation, tests, questionnaires, exercises and diaries have been used as data collection methods as well. 
A summary of the topics, number of participants and data collection methods is presented in Table \ref{tab:primaries}.

In addition to the literature review, a Google search on the topic was performed and one meaningful resource was found outside the scientific literature: LEGO for Software Engineering \cite{ccsu2021}.
However, the activities published on this website are presented in the papers mentioned and discussed above \cite{kurkovsky2019} \cite{kurkovsky2016} \cite{kurkovsky2015} \cite{kurkovsky2018}. 

Overall, despite an evidently positive application of LEGO in delivering Software Engineering concepts and practices to university students and IT professionals, there is a clear need to provide sufficient detail on the methods used as well as to produce empirical evidence that would support the claims presented by the researchers. 

\begin{table}
\renewcommand{\arraystretch}{1.3}
\caption{Summary of the primary papers}
\label{tab:primaries}
\centering
\begin{tabular}{c|p{2.5cm}|c|p{2cm}}
    \hline
    \textbf{ID} & \textbf{Topic} & \textbf{Participants} & \textbf{Data collection method} \\
    \hline
    \hline
    \cite{lynch2011} & Scrum & 43 & Survey \\
    \hline
    \cite{bica2020} & Scrum & 15 & Survey \\
    \hline
    \cite{steghofer2016} & Scrum & Not reported & Survey \\
    \hline
    \cite{paasivaara2014} & Scrum & 46 & Survey and lecture diary\\
    \hline
    \cite{lubke2005} & XP & 41 & Questionnaire \\
    \hline
    \cite{kurkovsky2016} & TDD & Not reported & Not reported \\
    \hline
    \cite{steghofer2019} & Software Processes & Not reported & Not reported \\
    \hline
    \cite{steghofer2018} & SPI & 20 & Survey \\
    \hline
    \cite{kurkovsky2015} & Requirements and Risk identification & 20 & Test \\
    \hline
    \cite{mayr2018} & Requirements & $\sim$200 & Survey \\
    \hline
    \cite{kurkovsky2019} & Requirements & 21 & Formal test \\
    \hline
    \cite{gama2019} & Requirements & 33 & Survey \\
     & Project Planning & 25 & Survey \\
    \hline
    \cite{hood2006} & Change Management & Not reported & Not reported \\
    \hline
    \cite{kurkovsky2018} & Construction & Not reported & Not reported \\
    \hline
    \cite{sablis2019} & GSE & 104 & Survey \\
    \hline
    \cite{laird2016} & Estimation techniques & 17 & In-class exercise \\
    \hline
\end{tabular}
\end{table}

\section{The Course: SENG403 -- Software Process and Product Quality}
\label{sec:seng403}

The SENG403 -- Software Process and Product Quality course introduces software quality key concepts, practices, methodologies and techniques present through the software lifecycle. 
SENG403 is a fourth-year elective course for Software Engineering and Computer Science (Hons.) students. It is also available as an elective for Master students in Information Systems and Information Technology.

SENG403 has been taught in the Software Engineering programme at the University of Canterbury since 2018.
The course has been designed by the author of the paper based on his experience in software industry and tertiary education since 2011.
The topics covered in the course include the following:
\begin{enumerate}
    \item \textbf{Quality basics.} An introduction to the basic concepts of quality, its relation to Software Engineering, the current obstacles and solutions for ensuring quality of software systems.
    \item \textbf{Quality models and standards.} A review of quality standards, conventions and models used in the industry.
    \item \textbf{Managing quality in software projects.} An overview of the quality management knowledge and techniques, including planning and control activities.
    \item \textbf{Quality in software processes.} An in-depth review of the challenges faced by organizations developing software products, and how those are managed through SPI initiatives.
    \item \textbf{Product quality.} A description of the quality approach focused on the quality attributes of a software product, the models, standards and validation and verification practices.
    \item \textbf{Introduction to measurement.} An overview to the need of measure, its basic concepts, the standards and methodologies related with metrics and measurement.
    \item \textbf{Tools and techniques}. A review of quality tools and techniques useful during management and assurance processes.
\end{enumerate}

At the end of SENG403, students are expected to understand, design and conduct, from a broader perspective, actions that improve the quality attributes of what/how they develop.
SENG403 has a total of nine learning outcomes (LOs), relevant for this paper being the following:
\begin{itemize}
    \item LO1 – Demonstrate advanced knowledge of quality concepts. Describe, discuss, and apply the principles of SQA.
    \item LO2 – Apply analysis skills to abstract and devise quality problems that affect process, product and people (the P's) in the software engineering context.
    \item LO4 – Evaluate complex and integral systems in order to recognize failures.
    \item LO6 – Compare and evaluate available solutions and apply the most suitable strategy to improve quality aspects of the P’s.
\end{itemize}

The full information of the course can be consulted at \cite{seng403}.

\section{KUALI-Brick Activity}
\label{sec:kuali-brick}
As part of SENG403, we carry out the KUALI-Brick activity towards the end of the course. The activity itself is described in detail in the following subsections.

\subsection{Objective and Target Audience}
The learning goal of the activity is to understand and apply the SQA and SPI concepts.
In order to achieve this, students need to: (i) understand, adopt and adapt a predefined building process; (ii) analyze and define the most relevant quality attributes to be assessed; and (iii) discuss and agree on the metrics to be collected. 

The purpose of the activity is to build a LEGO city within a fixed amount of time and obtaining as many points as possible. Points are allocated according to the quality assessment of the objects that the teams build.

The target audience of the activity consists of undergraduate
Software Engineering students enrolled in the fourth year of their Bachelor degree.

\subsection{Materials and Space}
\begin{figure}
    \centering
    \includegraphics[width=8cm]{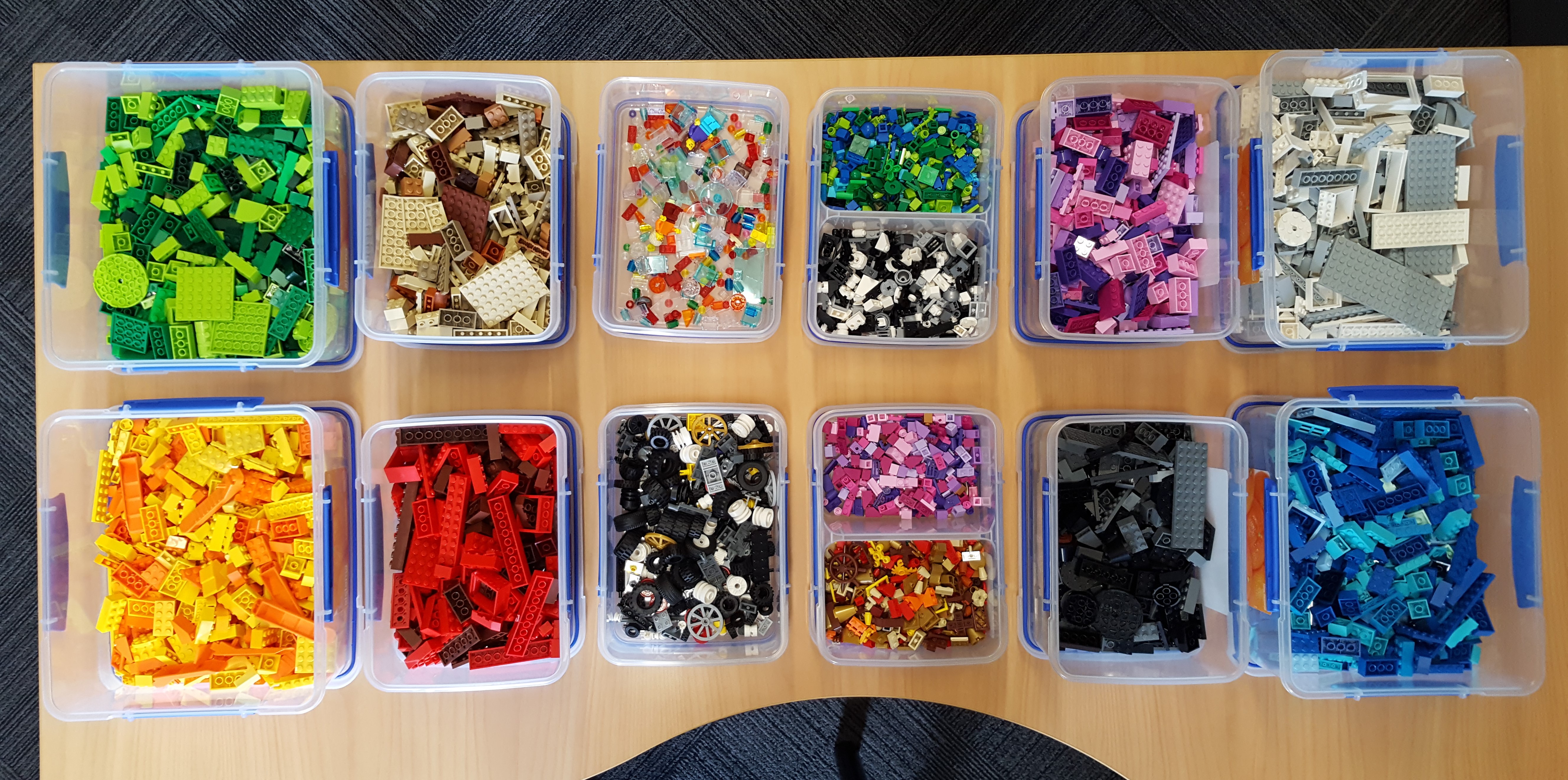}
    \caption{Boxes of bricks organized by colour and size.}
    \label{fig:brick-boxes}
\end{figure}

\begin{figure}
    \centering
    \includegraphics[width=8cm]{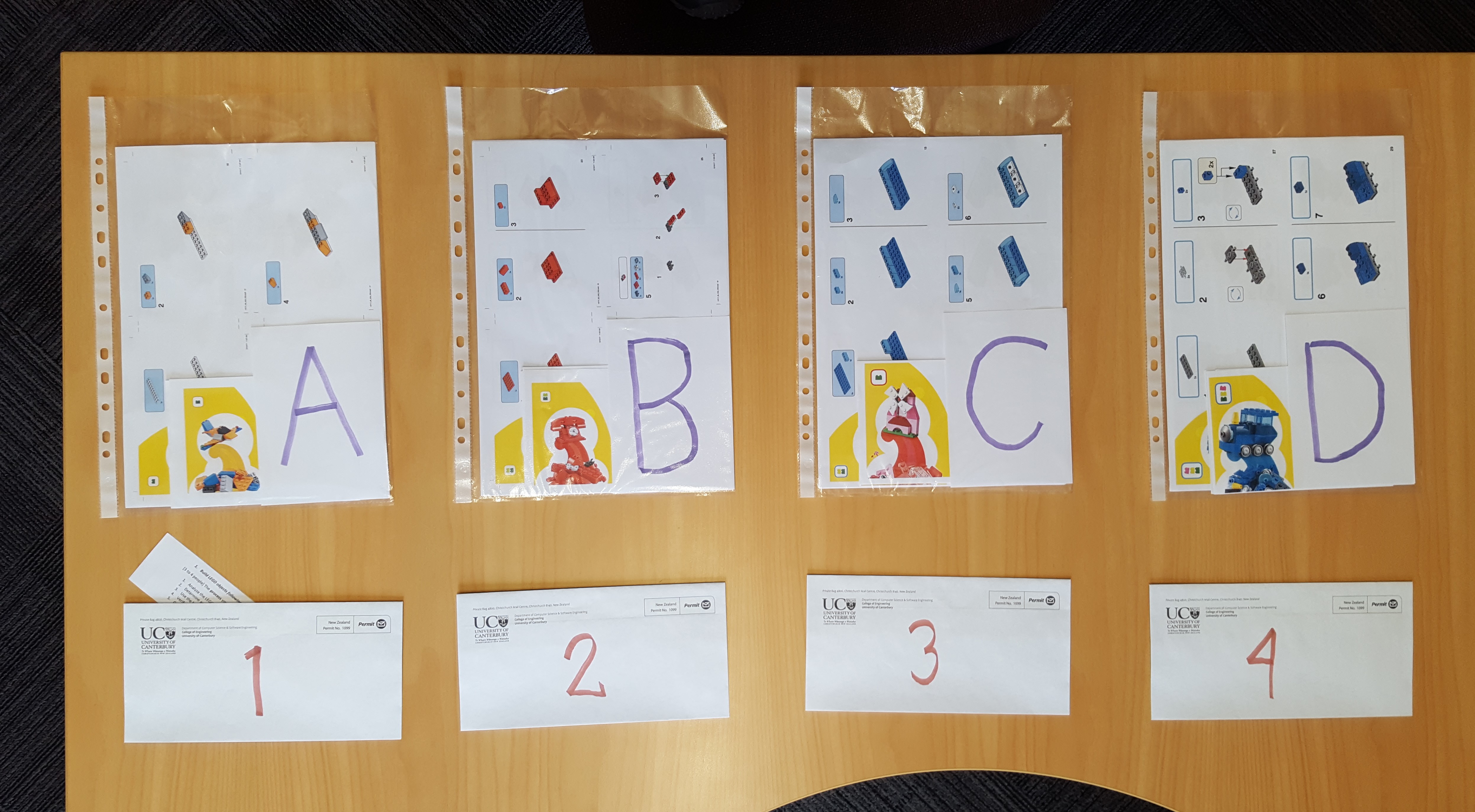}
    \caption{Envelopes and pockets containing the building processes and the objects.}
    \label{fig:envelopes-pockets}
\end{figure}

The participants have access to a wide variety of LEGO bricks, a predefined building process and a description of the expected objects to be built. 
The required material is the following:

\begin{itemize}
    \item 12 boxes of LEGO bricks (approx 3,000 bricks in total), the bricks are organized by colour and size. See Figure \ref{fig:brick-boxes}.
    \item 8 numbered envelopes containing building processes that each team will follow. There are three different building process ``approaches''. 
    \item 8 copy-safe pockets, each containing a set of four LEGO objects  that each team will have to build. The pockets are labeled with letters. See Figure \ref{fig:envelopes-pockets}.
    \item A city plan (optional).
    \item A card with the steps of the activity for each team (optional).
    \item A whiteboard, markers and a stopwatch.
    \item A projector (optional).
\end{itemize}

This material is sufficient for 32 students distributed in eight 4-member teams; however, the ideal number is 24 students distributed in six teams.

The space required for the activity is a flat room, ideally with a projector. A table per team is required with two bigger tables to sit the boxes with the bricks and the city plan, placed in the way to be easily accessible by the participants. 
The participants are distributed in teams around the room, while the bricks and the city plan are located in the middle.

\subsection{Activity Step-by-Step}
The KUALI-Brick session is designed to be carried out in 100 minutes and consists of three to four iterations. 
Three main stages can be identified: the Set-up and presentation stage is followed by the Building-and-Assessment iterations, with a wrap-up during the Discussion stage.

During the Set-up and presentation stage, the instructor sets up the materials and arranges the room, presents the generalities of the activity, explains the rules and divides the students into teams. 
The steps for this stage are listed below with suggested times for each step included in brackets:
\begin{enumerate}
    \setcounter{enumi}{-1}
    \item The instructor presents the learning goal and objective of the activity.
    \item The participants are asked to form teams of at most four people.
    \item As a team, they choose one of the copy-safe pockets labeled from A to H. It contains a set of LEGO objects that they will have to build.
    \item As a team, they discuss common quality attributes and their associated metrics that a LEGO build should have. Three to five attributes are usually enough \textbf{[5 minutes]}.
    \item As a group, they define and list quality attributes of a LEGO build, then choose three of them \textbf{[5 minutes]}.
    \item As a team, they assign points to each LEGO object they have. They have a budget of 30 points to allocate (allocated points must be integers). The students record them on the whiteboard \textbf{[7 minutes]}.
    \item They swap the set of objects with the team on their right hand \textbf{(Optional)}.
    \item As a team, they choose one of the envelopes numbered from 1 to 8. It contains the building process that each team will follow.
    \item As a team, they organize themselves according to the building process allocated to them \textbf{[3 minutes]}.
\end{enumerate}

During the Building-and-Assessment phase, the teams go through several iterations of building an object using the bricks and then assessing the objects built by other teams and allocating points to them based on predefined quality attributes. The steps for this stage are:

\begin{enumerate}
    \setcounter{enumi}{8}
    \item As a team and once the time starts counting, the students build the objects in the following order and one per iteration \textbf{[10 minutes]}: 
\begin{itemize}
    \item Animal
    \item Building
    \item Vehicle
    \item Object
\end{itemize}
    \item As a team and after each iteration, the students go to a different team and ask them to assess the quality of their LEGO build. \item As QAs, they assess the LEGO build using the attributes defined in 4. Then, based on their assessment, they assign points to the LEGO build \textbf{[3 minutes]}.
    \item As a team, the students get the points that the QA allocated to them and record them on the whiteboard.
    \item As a team, they place the LEGO object on the city plan.
    \item As a team, they discuss improvements to the building process and apply them in the next iteration \textbf{[2 minutes]}.
    \item The students repeat the steps until they have built all the LEGO objects assigned to the team.
\end{enumerate}

In the end, the Discussion stage is carried out as a group promoting reflection and a wrap-up of the session. 

\begin{enumerate}
    \setcounter{enumi}{15}
    \item As a group, the participants add the points up and discover the winner.
    \item They reflect on the activity and share their experience. 
Some questions to promote participation are:
\begin{itemize}
    \item Have you achieved your objective?
    \item Does your product have good quality?
    \item Did you come across any challenges when assessing the quality attributes?
    \item What are the advantages of the building process assigned to you?
    \item What are its disadvantages?
    \item How did you manage collaboration and communication issues, if any?
    \item How would you improve your building process?
\end{itemize}
\end{enumerate}

\begin{figure}
    \centering
    \includegraphics[width=8cm]{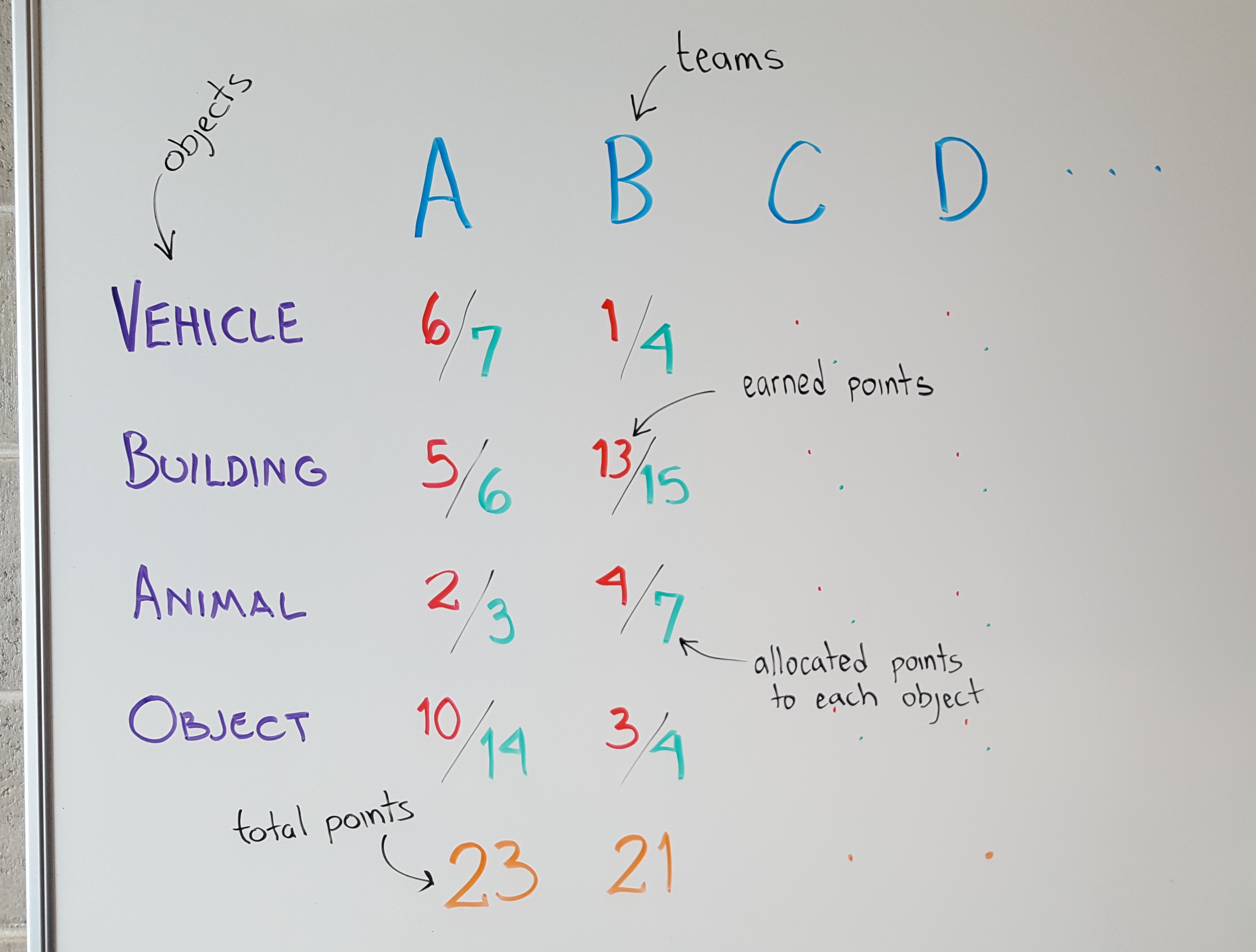}
    \caption{Allocated and earned points per object and team.}
    \label{fig:board}
\end{figure}

Figure \ref{fig:board} presents the board showing the teams (columns), objects (rows), points allocated to each object (green), points earned after the quality assessment is performed (red), and total points (orange) at the end of the activity.

\begin{figure}
    \centering
    \includegraphics[width=8cm]{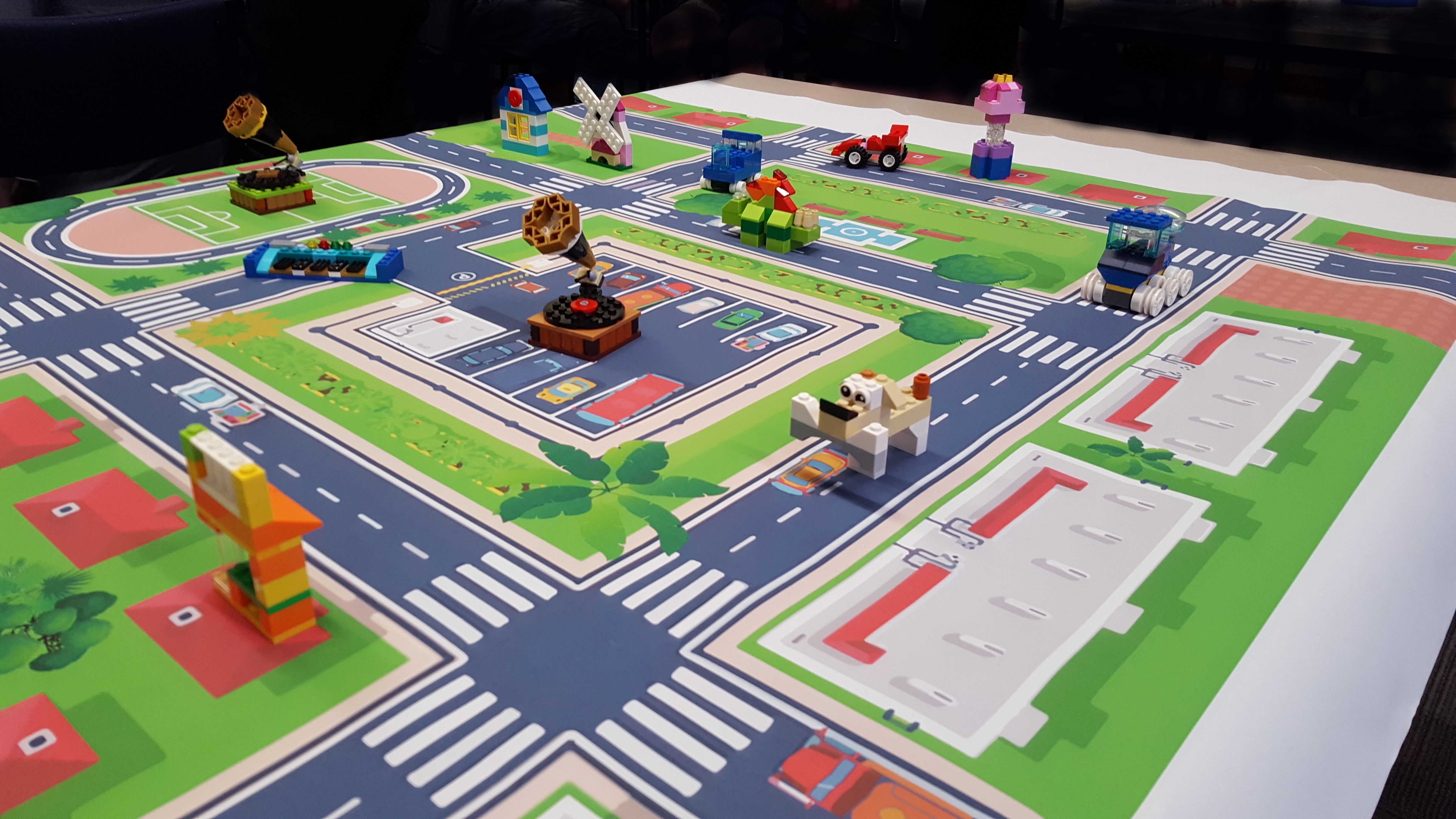}
    \caption{City plan at the end of the activity}
    \label{fig:city-plan}
\end{figure}

Figure \ref{fig:city-plan} shows the city plan once the last iteration is finished.

\subsection{Building Process Approaches}
Each team has to follow a predefined building process to create LEGO objects; there are three different approaches:

\textbf{Task-oriented} approach:
\begin{enumerate}
    \item Analyze the LEGO build requirements.
    \item Determine and select the necessary bricks.
    \item Use the bricks to build it.
    \item Verify if the build corresponds to the ones specified in the requirements.
    \item Get your build assessed.
\end{enumerate}

\textbf{Role-oriented} approach:
\begin{itemize}
    \item \textbf{Provider:} you have access to the LEGO specifications and you are in charge of selecting the required bricks. You give instructions to the Builder on how to assemble the object.
    \item \textbf{Builder:} you are in charge of building the object using the bricks and instructions given to you by Provider. You do not have access to the LEGO specifications.
    \item \textbf{Guide:} you are in charge of giving instructions to the Builder. You have to verify that the specified constraints are followed, and the object corresponds to the one from the specifications. You have access to the LEGO specifications. 
    \item \textbf{Assessor:} you are in charge of assessing the builds created by other teams.
\end{itemize}

\textbf{Organize yourselves} approach, which allows the team to decide on how they will run their process.

The teams must follow the approach at least during their first iteration, after which they are allowed to adjust and improve their building process for subsequent iterations. 
That is exactly the moment for the students to put their SPI knowledge in practice.

It is worth mentioning that more approaches can be added; however, in this activity, we decided to play with the rigidness of the process: offering options of the most rigid (task-oriented), a more balanced one (role-oriented), and the most flexible (self-organized).

Lastly, the quality aspect can be traced at different levels of the activity. 
While it is explicitly addressed as a role (Assessor) in the role-oriented approach, it becomes less visible in the task-oriented approach (step 4), and disappears in the self-organized approach.
This decision was made in order to provide students with a wider variety of situations, aiming at inspiring suggestions of improvement and awareness among students during the Discussion phase.

\subsection{SQA and SPI Come into the Game}

In order to understand how the activity connects theory and practice, a mapping between the SQA and SPI topics and the learning domains is described below using the cognitive levels of Bloom's taxonomy (L1-Remember, L2-Understand, L3-Apply, L4-Analyze, L5-Evaluate and L6-Create).

\textbf{Quality and Process definitions:}
\begin{itemize}
    \item \textit{Remember (L1)} the definition of quality, process and the components of a process (e.g. roles, tasks and products).
    \item \textit{Recognize (L1)} the process elements present (or absent) in their building process.
    \item \textit{Understand (L2)} the term quality in terms of the new context.  
\end{itemize}
Students were taught several definitions of quality (e.g. ISO/IEC 25010 \cite{isoiec25010} and IEEE 730 \cite{ieee730}) and process (e.g. ISO/IEC 12207 \cite{isoiec12207}).

\textbf{Process Improvement:}
\begin{itemize}
    \item \textit{Contrast (L2)} the advantages and disadvantages of each building process and its potential improvements.
    \item \textit{Implement (L3)} the improvements in a timely, effective and efficient manner.
    \item \textit{Detect (L5)} the nonconformities of their process and product.
\end{itemize}

As a strategy for guiding their SPI, students were taught the Plan–Do–Check–Act (PDCA) cycle, which is commonly used for continuous improvement of processes and products.

\textbf{Quality Assurance:}
\begin{itemize}
    \item \textit{Explain (L2)} the criteria to be used for the quality assessment of the products. 
    \item \textit{Execute (L3)} verification and validation procedures for checking that the product meets the specifications.
    \item \textit{Evaluate (L5)} the quality of the products based on objective criteria. 
    \item \textit{Critique (L5)} their own process and product in order to identify shortfalls and limitations.
\end{itemize}
During the course, students have been taught the SQA definitions (IEEE 610.12 \cite{ieee610} and IEEE 730) and process (CMMI-DEV \cite{sei2010} and ISO/IEC 12207).

\textbf{Quality models, attributes and metrics:} 
\begin{itemize}
    \item \textit{Recognize (L1)} the most relevant/valuable characteristics of the product.
    \item \textit{Analyze (L4)} the suitability of the quality models they know to the problem they are solving. 
    \item \textit{Create (L6)} a definition of their own quality attributes and metrics.
    \item \textit{Adapt (L6)} a known quality model to the current context and needs by providing adjusted quality attributes definitions.
\end{itemize}

Some examples of the quality attributes defined by the students are:
\begin{itemize}
    \item \textbf{Brick accuracy:} the degree to which the object conforms to the build requirements (color and shape of bricks).
    \item \textbf{Similarity:} the degree to which the object \textit{looks like} the picture in the specification. 
    \item \textbf{Playability/Entertainability:} the degree to which the object has attributes that makes it pleasing and satisfying to play with.
    \item \textbf{Structural integrity:} the degree to which the structure of the object is sound and resistant.
    \item \textbf{Building complexity:} the degree to which the object is easy to build.
\end{itemize}

To participate in the activity in the most advantageous manner, it is expected that the students will bear in mind and apply the ISO/IEC 25010 and the Goal Question Metric (GQM) approach as well as other knowledge previously discussed in class.

\subsection{Data Collection and Analysis}

The activity has been applied during three consecutive years with the following number of participants: 10 students in 2018, 27 in 2019 and 31 in 2020.
The participants possessed advanced knowledge about software engineering principles, practices and processes from other courses, and had gained industry experience in internships. 
Fifty-six students identified themselves as males while 12 as females.
Most of the students were from New Zealand (59) and the rest were overseas students from seven different nationalities.
Lastly, the students' age median was 22 and the average was 23.0. 

After the conclusion of the activity, the students were asked to answer a survey anonymously.
The survey consisted of two parts: the first part presented a subset of a standardized questionnaire developed by von Wangenheim et al. \cite{vonwangenheim2012}; in the second part the students were asked to choose from a list of nine LOs, which they believed were tackled by the activity, followed by their opinion as a free text.

Since 2018, 25 students have answered the survey. 
Table \ref{tab:survey_results} presents the average and median for each of the statements of the survey.

\begin{table}
\renewcommand{\arraystretch}{1.3}
  \caption{Survey results}
  \label{tab:survey_results}
\centering
\begin{tabular}{p{5cm}|c|c}
    \hline
    \textbf{Statement} & \textbf{Average} & \textbf{Median} \\
    \hline
    \hline
    The activity is attractive & 4.64 & 5 \\ 
    \hline
    The way the activity works suits my way of learning & 4.16 & 5 \\
    \hline
    The activity content is connected to other knowledge I already have & 4.36 & 4 \\        
    \hline
    While completing each step of the activity, I felt confident that I was learning & 3.76 & 4 \\
    \hline
    I had fun & 4.96 & 5 \\
    \hline
    I lost track of time during the activity & 4.72 & 5 \\
    \hline
    This activity is appropriately challenging for me, the tasks are neither too easy nor too difficult & 4.20 & 5 \\
    \hline
    I would recommend this activity to my peers & 4.72 & 5 \\
    \hline
    I would like to try similar activities in the future & 4.68 & 5 \\
    \hline
    The activity contributed to my learning & 4.28 & 5 \\    
    \hline
\end{tabular}
\end{table}

Based on the survey results and the students' opinion, the activity is highly rated in the fun factor (average: 4.96), students seem to enjoy and are engaged throughout the activity. They lose track of time (4.72) and would like to have similar activities in the future (4.68).

Although the activity was very well received, there are aspects on the learning side that could be improved, particularly in regards to emphasizing the learning aspects during the activity (3.76).

In the LOs part of the survey, 78\% of the participants perceived they achieved LO1, 67\% achieved LO4, with 56\% and 33\% choosing LO6 and LO2 respectively.
Two additional LOs were selected by the participants:
LO3 -- Hypothesize specific improvements to make the three P’s more effective, efficient and reliable, and LO5 -- Implement and justify quality management systems in a controlled and organized manner, allowing a continuous improvement. 
Both were selected by 44\% of the participants. 

Almost no difference in responses of the participants between individual years were observed; however, in 2019 the most number of students, out of those who participated in the activity, answered the survey (16 out of 25).

\subsection{Evolution of the Activity}
Over the years, the activity has evolved as based on the pursued learning goals and the target audience, at the same time improving in response to feedback from the participants.
The predecessor activity of KUALI-Brick was created in 2012; its motivation was to teach the basic concepts of Kanban (work-in-progress limit, board structure, work flow and team collaboration) to IT professionals differently.
The goal of the activity then was to build basic buildings (a school, a hospital, a two-level house, etc.) using bricks, slope bricks and plates (1x1, 2x1, 2x2 and 2x4). 
In addition to the buildings, the participants had to draw a city plan in an A0 paper sheet using markers.
The main condition was that teams had to coordinate their work and communicate through the Kanban board only.

Given the limited variety of bricks and being process- rather than product-focused, the buildings were simple and not so challenging.
A couple of years later, the buildings were replaced by sets of `2D objects' (e.g. a ghost, a banana and a watermelon in one set), which increased the challenge level resulting in more fun and engagement for the participants. 

In the following few years, the focus of the activity changed from Kanban to SPI, where the participants discussed improvements and adapted their process and workflow after each iteration.
With the aim of making the activity more competitive, another step was added, in which teams had to `sell' their builds to other teams in order to gain money.
In the end, this `selling' step was formalized as a quality assessment and SQA concepts were added into the mix.
Since the focus shifted from Kanban to SPI and SQA, the Kanban board was removed, the bricks variety increased and the types of builds became more complex.

\section{Discussion}
\label{sec:discussion}
The discussion is organized in four subsections: positive aspects, negative aspects, lessons learned and limitations of the study.

\subsection{Positive Aspects}
As reported by most of the LEGO-related studies presented in Section \ref{sec:related_work}, participants experience an increase in engagement and enjoyment.

We have observed an overwhelming evidence that the fun level of the students rockets as fun has obtained the highest score (4.96) in the survey.
In addition, students have commented that the activity was the \textit{``most fun lecture I've had in 3.5 years''}, \textit{``loved the activity, thank you for putting it together''}.

The engagement of the participants is very high: \textit{``it was really engaging, and got us to think about to topic in a context other than software which I think allows for different perspectives''}.

The activity breaks the routine and makes the time more enjoyable:
\textit{``this was one of the best activities I've done at uni''}. 
The playful nature of the activity positively influences the attitude of the participants towards the learning process: \textit{``it was a fun and memorable way of learning the concepts''}.

Its hands-on approach gives the opportunity to put theoretical concepts into practice: \textit{``I thought it was a fantastic way of putting into practice some of the concepts we've learned in class''}. 

The fact that students, being used to construct `abstract' software, can see the palpable objects they have build, facilitates acquisition of such abstract concepts as quality, and their transformation into concrete applications. 
\textit{``It helped apply some of the concepts that I had learned from the course into a game like scenario which is more of a concrete structure then the abstract nature of software''}.
As a consequence, the participants realize how a quality attribute or metric affects the assessment process of a real product. 
They also become aware of how a timely adjustment can improve the building process by making it more efficient or by enhancing the quality of the final outcome.

\subsection{Negative Aspects}
Stress associated with short iterations \cite{steghofer2018} and time constraints have been observed in other studies. 
In this experience, time was also a topic of concern. 
Not having enough time for building the objects and/or discussing at the end of the activity were the main issues expressed by participants.
\textit{``Due to the overrun in time restricting the opportunity for reflection and discussion about the relation to software process and quality''}.
It was observed that the lack of a sufficient discussion time creates difficulty in connecting theory and practice.

As mentioned before, the set-up of the room is quite important. In this experience, in 2018 the boxes with bricks were sitting on a table on one side of the room, giving participants only two sides of the table to reach the bricks. 
For subsequent instances, the boxes were located in a table in the middle of the room, drastically improving access to them: \textit{``A few things just need to be ironed out like the mosh pit trying to get lego and the timings of things''}. 

Another constraint might be the price of the game material; however, the output is worth the investment. The cost of the bricks, boxes and the city plan printing was approximately \$1,000 NZD (\$715 USD).

\subsection{Lessons Learned}
\textbf{The discussion at the end of the activity is fundamental in order to emphasize the learning outcomes.} 
Participants lose track of time easily and at early stages of the activity. In addition, the time allocated is tight, therefore it is most likely that the discussion at the end of the activity might be affected due to time constraints. 
In the 2019 instance of the activity, the discussion was shortened from 20 to 5 minutes, causing disagreement expressed by some students.
On the contrary, when the discussion is carried out as planned, the learning aspects are boosted -- demonstrated by a comment from the 2020 instance: \textit{``The discussion guided by Miguel about how the activity was *really* valuable and it really helps to relate what we did to what we have been learning''}.

\textbf{Prepare students for the activity, let them know what will happen.} 
For the 2019 instance, students were informed about the activity with no major details provided, which was supposed to be a `surprise'. 
It might not be an issue due to the fact that the activity is neither formally assessed nor requires preparation from students; however, some students can perceive it differently: \textit{``it is very different to normal activities undertaken in class and was somewhat of a shock when I had come prepared to take notes and the like''}. 
In order to avoid undesirable surprises, we included the activity into the course schedule and it is advisable to remind students of the activity once again.

\textbf{Make the most of the fun boost.}
In order to make the most of the activity, we suggest conducting it as an optional activity of the course; this motivates the students to have the right attitude.
The impact of the activity is important, keeping the momentum going in the participants long afterwards and outside the course content; this was observed from 2019 and 2020 students: \textit{``It was so fun, we definitely needed an activity like with all the work we've been putting towards the end of the semester with all the deadlines around us''}; and \textit{``It was certainly a welcome activity to have some fun during a very stressful period of time. Please keep running this!''}.

\textbf{Aesthetics are important.} 
Such aspects as the set-up of the room, the bricks in boxes organized by colour and size, and the city plan have a positive effect on students. 
From the beginning the students enter the room, the experience should be welcoming and making them feel that the lecture material was specially prepared for them -- this breaks the routine and boosts engagement, which is transferable to following lectures.

\textbf{Custom vs. Packaged LEGO sets.} 
On the one hand, custom sets are more difficult to build and assess as they introduce extra variables into the activity making it more complex. For example, participants have to rely on their creativity and ability to build, so the time becomes an issue.
Packaged sets, on the other hand, offer a clear image of what the final product should look like, at the same time reducing the creativity part.
We suggest avoiding branded sets (e.g. Star Wars or Friends) and using the LEGO Classics ideas that contain small sets and ideas to build with no limits to creativity. Up to now, the Classics set has brought the better results in our experience. 

\textbf{Add variability to the final iteration.} 
After the second iteration participants might perceive the dynamic as repetitive, therefore we suggest addition of some variability to the third iteration.
The simplest and most effective variation in our experience was asking the teams to build two instead of one objects during the third iteration. With the time constraints in place, the participants are forced to rush their process causing the quality of the products to be affected. 
We call this variation ``On a rush!''. 
Another suggestion is to shuffle team members around the teams in order to make the process adjustments more challenging.

\subsection{Limitations}
The main limitation of this work is the lack of a formal assessment to confirm improvement in the participants' knowledge.
Currently, the claim of improvement is based solely on observations and participants' opinions.

Secondly, while the activity allows the students to experience, relate and reflect on quality attributes, metrics, SQA and SPI issues, they are exposed to simplified versions of those concepts only. 
As stated by \cite{sablis2019}, the LEGO metaphor allows to simulate some characteristics of software development, but it is not a substitute for actual software development tasks.

Lastly, if the playful aspect is not controlled adequately, the activity can turn into a chaos, given the euphoria of participants and the relaxed environment in which it is carried out. 
Here, the instructor's experience plays an important role in controlling the dynamics of the class and in keeping the participants focused on the learning goals.

\section{Conclusions and Future Work}
\label{sec:conclusions}

While knowledge acquisition is one of the most important aspects in every learning context, there are other relevant aspects that help creating an effective learning environment in a lecture room.
The attitude towards the lecturer, other classmates and the course content are variables that affect the student performance and achievement of the learning outcomes.
The fun aspect is an unexpected attribute during a fourth-year Engineering lecture; however, adding it to the mix can affect the student attitude in a positive way.

The KUALI-Brick activity was created with the aim of involving students into application of theoretical concepts and giving them the opportunity to experience those concepts firsthand.
We can conclude that the KUALI-Brick activity resulted in an excellent tool for making fourth-year Software Engineering students aware of how abstract concepts can be instantiated in concrete objects.

In general, we can state that the research around the combination of LEGO and education in Software Engineering is promising. This mix almost secures a superb learning experience in terms of fun, engagement and enjoyment, but still needs evidence to become recognized as a game that delivers quantitative improvement in the learning outcomes of the participants. 

As future work, an assessment item will be created to evaluate the learning aspect of the activity. Besides, the KUALI-Brick activity rules and steps will be made available in order to share and obtain feedback from the community.

\section*{Acknowledgment}
The author would like to thank Polina Maslova for her contribution to the activity design.

\bibliographystyle{IEEEtran}
\bibliography{KUALI-Brick.bib}

\end{document}